\documentclass[10pt,twocolumn]{usetex-v1}

\usepackage{times}
\usepackage{graphicx}
\usepackage{subfigure}
\usepackage{epstopdf}
\usepackage{amsfonts}
\usepackage{amssymb}
\usepackage{amsmath}
\usepackage{makecell}
\usepackage{dsfont}
\usepackage{threeparttable}
\usepackage{pifont}
\usepackage{array}
\usepackage{multirow}
\usepackage{cite}
\usepackage{caption}
\usepackage{endnotes}
\usepackage[hyphens]{url}
\usepackage{enumerate}
\usepackage{booktabs}

\renewcommand{\paragraph}[1]{\noindent {\bf #1}}
\begin{document}
	
\title{CDStore: Toward Reliable, Secure, and Cost-Efficient Cloud
Storage via Convergent Dispersal}

\author{Mingqiang Li\thanks{Mingqiang Li is now with Hong Kong Advanced
Technology Center, Ecosystem \& Cloud Service Group, Lenovo Group Ltd. This
work was done when he was with the Chinese University of Hong Kong.} , Chuan
Qin, and Patrick P. C. Lee\\
{\em Department of Computer Science and Engineering, The Chinese
University of Hong Kong}\\
{\em mingqiangli.cn@gmail.com, \{cqin,pclee\}@cse.cuhk.edu.hk}}

\maketitle

\begin{abstract}
We present {\em CDStore}, which disperses users' backup data
across multiple clouds and provides a unified multi-cloud storage solution
with reliability, security, and cost-efficiency guarantees.  CDStore builds on
an augmented secret sharing scheme called {\em convergent dispersal}, which
supports deduplication by using deterministic content-derived hashes as inputs
to secret sharing.  We present the design of CDStore, and in
particular, describe how it combines convergent dispersal with two-stage
deduplication to achieve both bandwidth and storage savings and be robust
against side-channel attacks.  We evaluate the performance of our
CDStore prototype using real-world workloads on LAN and commercial cloud
testbeds.  Our cost analysis also demonstrates that CDStore achieves a
monetary cost saving of 70\% over a baseline cloud storage solution using
state-of-the-art secret sharing. 
\end{abstract}

\section{Introduction}

Cloud storage provides cost-efficient means for organizations to host backups
off-site \cite{Zdnet14}.  However, from users' perspectives, putting all data
in one cloud raises {\em reliability} concerns regarding the single point of
failure \cite{Armbrust10} and vendor lock-in \cite{Abu-Libdeh10}, 
especially when cloud storage providers can spontaneously terminate their
business \cite{Nirvanix}. Cloud storage also raises {\em security} concerns,
since data management is now outsourced to third parties.  Users often want
their outsourced data to be protected with guarantees of {\em confidentiality}
(i.e., data is kept secret from unauthorized parties) and {\em integrity}
(i.e., data is uncorrupted). 

{\em Multi-cloud storage} coalesces multiple public cloud storage services
into a single storage pool, and provides a plausible way to realize both
reliability and security in outsourced storage. 
It disperses data with some form of redundancy across multiple clouds,
operated by independent vendors, such that the stored data can be recovered
from a subset of clouds even if the remaining clouds are unavailable.
Redundancy can be realized through {\em erasure coding} (e.g., Reed-Solomon
codes \cite{Reed60}) or {\em secret sharing} (e.g., Shamir's scheme
\cite{Shamir79}).  Recent multi-cloud storage systems (e.g.,
\cite{Kotla07,Bowers09,Abu-Libdeh10,Hu12,Wu13}) leverage erasure coding to
tolerate cloud failures, but do not address security;
DepSky \cite{Bessani13} uses secret sharing to
further achieve both reliability and security.  Secret sharing
often comes with high redundancy, yet its variants are shown to reduce
the redundancy of secret sharing to be slightly higher than that of erasure
coding, while achieving security in the computational sense (see
\S\ref{sec:background}).  
Secret sharing has a side benefit of providing {\em keyless} security (i.e.,
eliminating encryption keys), which builds on the difficulty for an
attacker to compromise multiple cloud services rather than a secret key.
This removes the key management overhead as found in key-based encryption
\cite{Storer09}. 

However, existing secret sharing algorithms prohibit storage savings achieved
by {\em deduplication}.  Since backup data carries substantial identical
content \cite{Wallace12}, organizations often use deduplication to save
storage costs, by keeping only one physical data copy and having it shared by
other copies with identical content.  On the other hand, secret sharing uses
random pieces as inputs when generating dispersed data.  Users embed different
random pieces, making the dispersed data different even if the original data
is identical.  

This paper presents a new multi-cloud storage system
called {\em CDStore}, which makes the first attempt to 
provide a unified cloud storage solution with reliability, security, and
cost efficiency guarantees.  CDStore builds on our prior proposal of an
enhanced secret sharing scheme called {\em convergent dispersal} \cite{Li14b},
whose core idea is to replace the random inputs of traditional secret sharing
with deterministic cryptographic hashes derived from the original data, while
the hashes cannot be inferred by attackers without knowing the whole original
data.   
This allows deduplication, while preserving the
reliability and keyless security features of secret sharing.  
Using convergent dispersal, CDStore offsets dispersal-level redundancy due to
secret sharing by removing content-level redundancy via deduplication, and
hence achieves cost efficiency.  
To summarize, we extend our prior work \cite{Li14b} and make three new
contributions.

First, we propose a new instantiation of convergent dispersal called 
{\em CAONT-RS}, which builds on AONT-RS \cite{Resch11}.  CAONT-RS maintains
the properties of AONT-RS, and makes two enhancements: (i) using OAEP-based
AONT \cite{Boyko99} to improve performance and (ii) replacing random inputs
with deterministic hashes to allow deduplication.  Our evaluation also shows
that CAONT-RS generates dispersed data faster than our prior AONT-RS-based
instantiation \cite{Li14b}. 

Second, we present the design and implementation of CDStore.  It
adopts {\em two-stage deduplication}, which first deduplicates data of the
same user on the client side to save upload bandwidth, and then deduplicates
data of different users on the server side to further save storage.  Two-stage
deduplication works seamlessly with convergent dispersal, achieves bandwidth
and storage savings, and is robust against side-channel attacks
\cite{Harnik10,Halevi11}.  We also carefully implement CDStore to mitigate
computation and I/O bottlenecks. 

Finally, we thoroughly evaluate our CDStore prototype using both
microbenchmarks and trace-driven experiments. We use real-world backup and
virtual image workloads, and conduct evaluation on both LAN and commercial
cloud testbeds.  We show that CAONT-RS encoding achieves around 180MB/s with
only two-thread parallelization.  We also identify the bottlenecks when
CDStore is deployed in a networked environment.  Furthermore, we show via cost
analysis that CDStore can achieve a monetary cost saving of 70\% via
deduplication over AONT-RS-based cloud storage.

\section{Secret Sharing Algorithms}
\label{sec:background}

We conduct a study of the state-of-the-art secret sharing algorithms.
A secret sharing algorithm operates by transforming a data input called
{\em secret} into a set of coded outputs called {\em shares}, with the
primary goal of providing both fault tolerance and confidentiality guarantees
for the secret.  Formally, a secret sharing algorithm is defined based on
three parameters $(n,k,r)$: {\em an $(n,k,r)$ secret sharing
algorithm (where $n > k > r \geq 0$) disperses a secret into $n$ shares such
that (i) the secret can be reconstructed from any $k$ shares, and (ii) the
secret cannot be inferred (even partially) from any $r$ shares.}

The parameters $(n,k,r)$ define the protection strength of a secret sharing
algorithm.  Specifically, $n$ and $k$ determine the fault tolerance degree of
a secret, such that the secret remains available as long as any $k$ out of $n$
shares are accessible. In other words, it can tolerate the loss of $n-k$
shares.  The parameter $r$ determines the confidentiality degree of a secret,
such that the secret remains confidential as long as no more than $r$ shares
are compromised by an attacker.  On the other hand, a secret sharing algorithm
makes the trade-off of incurring additional storage.  We define the 
{\em storage blowup} as the ratio of the total size of $n$ shares to the size
of the original secret.  Note that the storage blowup must be at least
$\frac{n}{k}$, as the secret is recoverable from any $k$ out of $n$ shares.

\begin{table}[t]
  \centering
  \small
  \begin{threeparttable}
  \begin{tabular}{lp{1.1in}p{0.8in}}
    \toprule
    \textbf{Algorithm} & \textbf{Confidentiality degree} & \textbf{Storage blowup\tnote{$\dag$}} \\
    \midrule
    SSSS \cite{Shamir79} & $r=k-1$ & $n$ \\
    IDA \cite{Rabin89} & $r=0$ & $\frac{n}{k}$ \\
    RSSS \cite{Blakley85} & $r\in[0,k-1]$ & $\frac{n}{k-r}$ \\
    SSMS \cite{Krawczyk93} & $r=k-1$ & $\frac{n}{k} + n \cdot \frac{S_{key}}{S_{sec}}$ \\
    AONT-RS \cite{Resch11} & $r=k-1$ & $\frac{n}{k} + \frac{n}{k} \cdot \frac{S_{key}}{S_{sec}}$ \\
    \bottomrule
  \end{tabular}
  \begin{tablenotes}
  \footnotesize
  \item[$\dag$]$S_{sec}$: size of a secret; $S_{key}$: size of a random
  key.
  \end{tablenotes}
  \end{threeparttable}
  \vspace{-6pt}
  \caption{Comparison of secret sharing algorithms.}
  \label{tab:dispersal}
  \vspace{-6pt}
\end{table}

Several secret sharing algorithms have been proposed in the literature.
Table~\ref{tab:dispersal} compares them in terms of the confidentiality degree
and the storage blowup, subject to the same $n$ and $k$.  Two extremes of
secret sharing algorithms are Shamir's secret sharing scheme (SSSS)
\cite{Shamir79} and Rabin's information dispersal algorithm (IDA)
\cite{Rabin89}.  SSSS achieves the highest confidentiality degree (i.e.,
$r=k-1$), but its storage blowup is $n$ (same as replication).  IDA
has the lowest storage blowup $\frac{n}{k}$, but its confidentiality
degree is the weakest (i.e., $r=0$), and any share can reveal the information
of the secret.  Ramp secret sharing scheme (RSSS) \cite{Blakley85} generalizes
both IDA and SSSS to make a trade-off between the confidentiality degree and
the storage blowup.  It evenly divides a secret into $k-r$ pieces, and
generates $r$ additional random pieces of the same size.  It then transforms
the $k$ pieces into $n$ shares using IDA.

Secret sharing made short (SSMS) \cite{Krawczyk93} combines IDA and SSSS using
traditional key-based encryption.  It first encrypts the secret with a random
key and then disperses the encrypted secret and the key using IDA and SSSS,
respectively.  Its storage blowup is slightly higher than that of IDA,
while it has the highest confidentiality degree $r=k-1$ as in SSSS.  Note that
the confidentiality degree is defined in the computational sense, that is, it
is computationally infeasible to break the encryption algorithm without
knowing the key.

AONT-RS \cite{Resch11} further reduces the storage blowup of SSMS, while
preserving the highest confidentiality degree $r=k-1$ (in the computational
sense).  It combines Rivest's all-or-nothing transform (AONT) \cite{Rivest97}
for confidentiality and Reed-Solomon coding \cite{Reed60,Blomer95} for
fault tolerance.  It first transforms the secret into an AONT package with a
random key, such that an attacker cannot infer anything about the AONT package
unless the whole package is obtained.  Specifically, it splits a secret into a
number $s \ge 1$ of words, and adds an extra \emph{canary} word for integrity
checking.  It masks each of the $s$ words by XOR'ing it with an index value
encrypted by a random key.  The $s$ masked words are placed at the start of
an AONT package.  One more word, obtained by XOR'ing the same random key with
the hash of the masked words, is added to the end of the AONT package.  The
final AONT package is then divided into $k$ equal-size shares, which are
encoded into $n$ shares using a {\em systematic} Reed-Solomon code
(a systematic code means that the $n$ shares include the original $k$ shares).

The security of existing secret sharing algorithms lies in the embedded
random inputs (e.g., a random key in AONT-RS).  Due to randomness, secrets
with identical content lead to distinct sets of shares, thereby prohibiting
deduplication.   This motivates CDStore, which enables secret
sharing with deduplication. 

\section{CDStore Design}
\label{sec:overview}

\begin{figure}[t]
\centering
\includegraphics[width=0.95\linewidth]{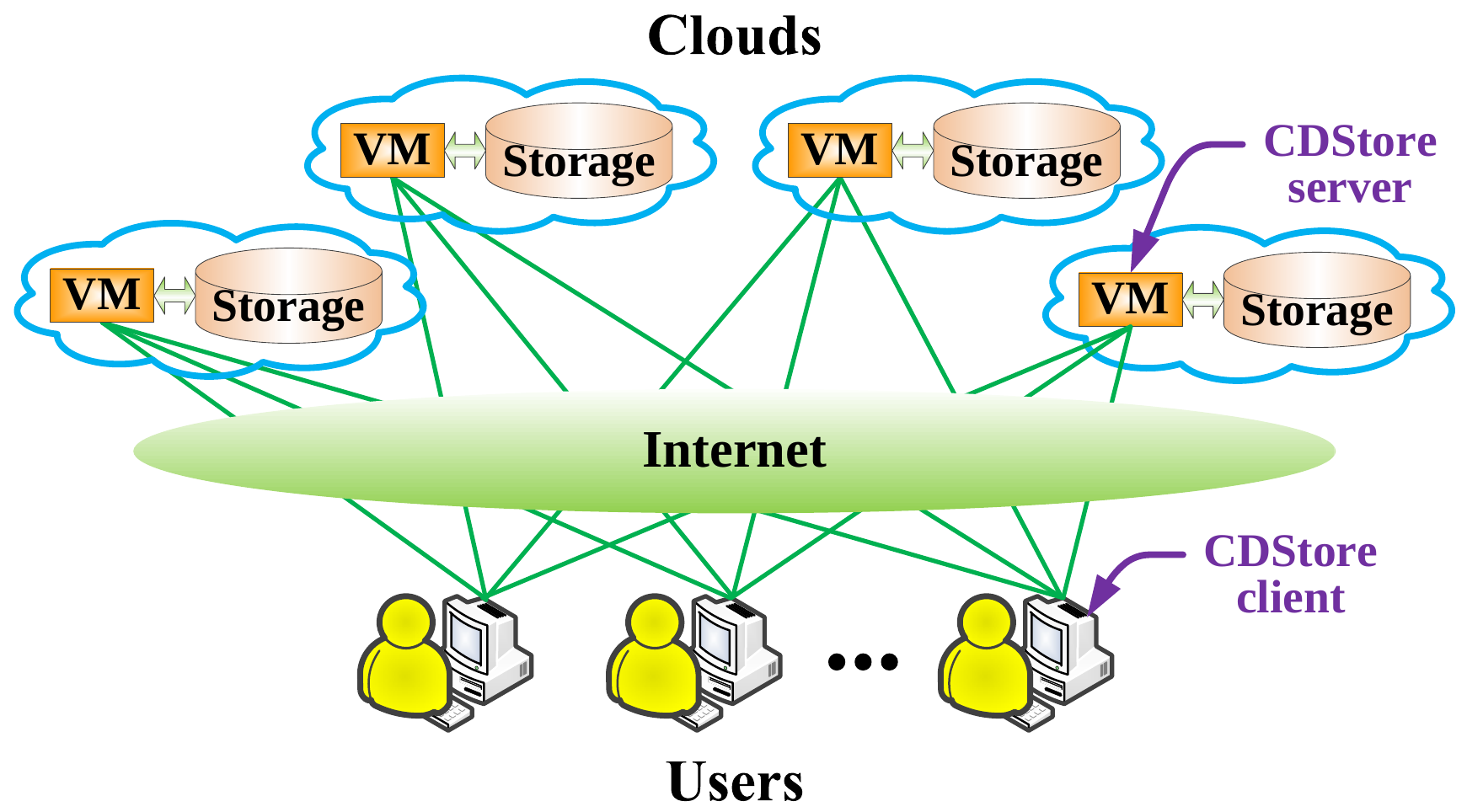}
\vspace{-6pt}
\caption{CDStore architecture.}
\label{fig:arch}
\vspace{-6pt}
\end{figure}

CDStore is designed for an organization to outsource the storage
of data of a large group of users to multiple cloud vendors.  It builds on the
client-server architecture, as shown in Figure~\ref{fig:arch}.  Each user of
the same organization runs the {\em CDStore client} to store and access its
data in multiple clouds over the Internet.  In each cloud, a co-locating virtual
machine (VM) instance owned by the organization runs the {\em CDStore server}
between multiple CDStore clients and the cloud storage backend.

CDStore targets backup workloads.  We consider a type of backups obtained by
snapshotting some applications, file systems, or virtual disk images.
Backups generally have significant identical content, and this makes
deduplication useful.
Field measurements on backup workloads show that deduplication can reduce the
storage overhead by 10$\times$ on average, and up to 50$\times$ in some cases
\cite{Wallace12}.   In CDStore deployment, each user machine submits a series
of backup files (e.g., in UNIX {\tt tar} format) to the co-located CDStore
client, which then processes the backups and uploads them to all clouds.

\subsection{Goals and Assumptions}
\label{subsec:goal}

We state the design goals and assumptions of CDStore in three aspects:
reliability, security, and cost efficiency.


\paragraph{Reliability:} CDStore tolerates failures of cloud storage providers
and even CDStore servers.  Outsourced data is accessible if a tolerable number
of clouds (and their co-locating CDStore servers) are operational.  CDStore
also tolerates client-side failures by offloading metadata management to the
server side (see \S\ref{subsec:CDStore-client-metadata}).  In the presence of
cloud failures, CDStore reconstructs original secrets and then rebuilds the
lost shares as in Reed-Solomon codes \cite{Reed60}.  We do not consider
cost-efficient repair \cite{Hu12}. 


\paragraph{Security:} CDStore 
exploits multi-cloud diversity to ensure {\em confidentiality} and {\em
integrity} of outsourced data against outsider attacks, as long as a tolerable
number of clouds are uncompromised.  Note that the confidentiality guarantee
requires that the secrets be drawn from a very large message space, so that
brute-force attacks are infeasible \cite{Bellare13a}.  CDStore also uses
two-stage deduplication (see \S\ref{subsec:twostage}) to avoid insider
side-channel attacks \cite{Harnik10,Halevi11} launched by malicious users.  
Here, we do not consider strong attack models, such as Byzantine faults in
cloud services \cite{Bessani13}.  We also assume that the client-server
communication over the network is protected, so that an attacker cannot infer
the secrets by eavesdropping the transmitted shares. 


\paragraph{Cost efficiency:} CDStore uses deduplication to reduce both
bandwidth and storage costs. 
It also incurs limited overhead in computation (e.g., VM usage)
and storage (e.g., metadata).  We assume that there is no billing for the
communication between a co-locating VM and the storage backend
of the same cloud, based on today's pricing models of most cloud vendors
\cite{Jellinek14}.





\subsection{Convergent Dispersal}
\label{subsec:CD}

\begin{figure*}[t]
\centering
\begin{tabular}{c@{\hspace{0.2in}}c}
\parbox[t]{2.1in}{
\centering
\includegraphics[height=1.1in]{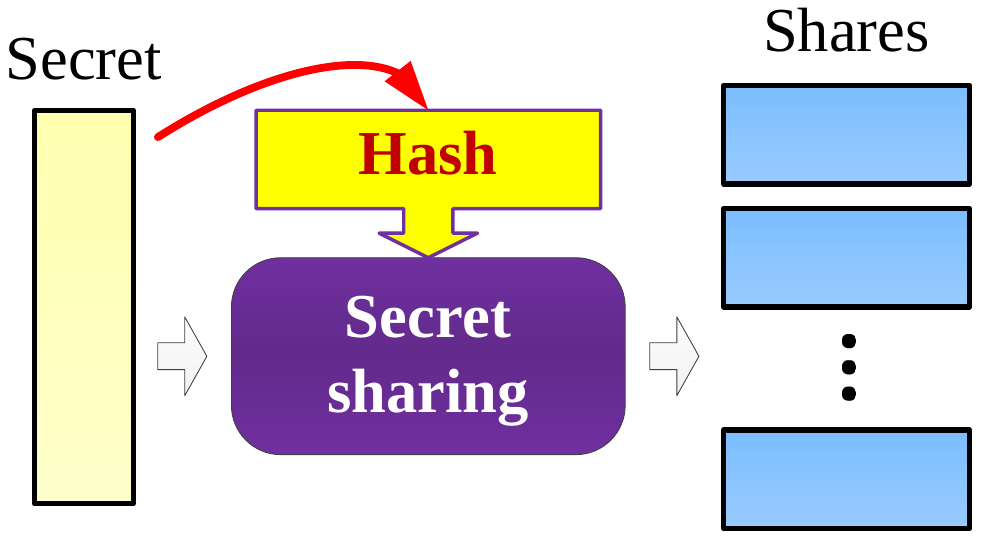}
\caption{Idea of convergent dispersal.}
\label{fig:main_idea}
}
&
\parbox[t]{4in}{
\centering
\includegraphics[height=1.2in]{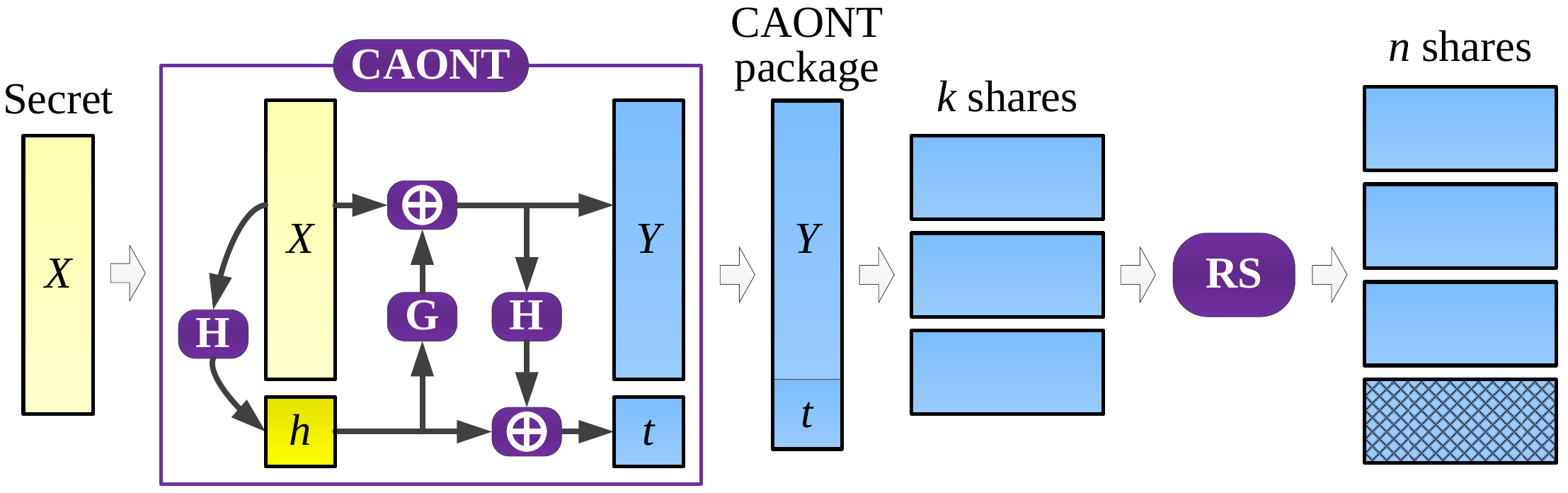}
\caption{Example of CAONT-RS with $n=4$ and $k=3$.}
\label{fig:convergent-dispersal}
}
\end{tabular}
\end{figure*}

Convergent dispersal enables secret sharing with deduplication
by replacing the embedded random input with a deterministic
cryptographic hash derived from the secret.  Thus, two secrets with
identical content must generate identical shares, making deduplication
possible.  Also, it is computationally infeasible to infer the hash without
knowing the whole secret.  Our idea is inspired by convergent encryption
\cite{Douceur02} used in traditional key-based encryption, in which the random
key is replaced by the cryptographic hash of the data to be encrypted.
Figure~\ref{fig:main_idea} shows the main idea of how we augment a secret
sharing algorithm with convergent dispersal. 


This paper proposes a new instantiation of convergent dispersal 
called {\em CAONT-RS}, which inherits
the reliability and security properties of the original AONT-RS, and makes two 
key modifications.  First, to improve performance, CAONT-RS replaces Rivest's
AONT \cite{Rivest97} with another AONT based on \emph{optimal asymmetric
encryption padding (OAEP)} \cite{Bellare94,Boyko99}.  The rationale is that
Rivest's AONT performs multiple encryptions on small-size words (see
\S\ref{sec:background}), while OAEP-based AONT performs a single encryption on
a large-size, constant-value block.  Also, OAEP-based AONT provably
provides no worse security than any AONT scheme \cite{Boyko99}.  Second,
CAONT-RS replaces the random key in AONT with a deterministic cryptographic
hash derived from the secret.  Thus, it preserves content similarity in
dispersed shares and allows deduplication.  Our prior work \cite{Li14b}
also proposes instantiations for RSSS \cite{Blakley85} and AONT-RS (based on
Rivest's AONT) \cite{Resch11}.  Our new CAONT-RS shows faster
encoding performance than our prior AONT-RS-based instantiation 
(see \S\ref{subsec:encoding-speed}). 


We now elaborate on the encoding and decoding of CAONT-RS, both of which are
performed by a CDStore client.  Figure~\ref{fig:convergent-dispersal} shows an
example of CAONT-RS with $n=4$ and $k=3$ (and hence $r=k-1=2$).


\paragraph{Encoding:}
We first transform a given secret $X$ into a CAONT package. Specifically, we
first generate a hash key $h$, instead of a random key, derived from $X$ using
a (optionally salted) hash function $\mathbf{H}$ (e.g., SHA-256):
\begin{equation}\label{equ:hash-func}
    h=\mathbf{H}(X).
\end{equation}
To achieve confidentiality, we transform $(X,h)$ into a CAONT package $(Y,t)$
using OAEP-based AONT, where $Y$ and $t$ are the head and tail parts of the
CAONT package and have the same size as $X$ and $h$, respectively.  To
elaborate, $Y$ is generated by:
\begin{equation}\label{equ:main-part-calulation}
    Y= X \oplus \mathbf{G}(h),
\end{equation}
where `$\oplus$' is the XOR operator and $\mathbf{G}$ is a generator function
that takes $h$ as input and constructs a mask block with the same size as $X$.
Here, we implement the generator $\mathbf{G}$ as:
%
\begin{equation}\label{equ:generator}
    \mathbf{G}(h) = \mathbf{E}(h,C),
\end{equation}
where $C$ is a constant-value block with the same size as $X$, and
$\mathbf{E}$ is an encryption function (e.g., AES-256) that encrypts $C$ using
$h$ as the encryption key.

The tail part $t$ is generated by:
\begin{equation}\label{equ:tail-part-calulation}
    t = h \oplus \mathbf{H}(Y).
\end{equation}


Finally, we divide the CAONT package into $k$ equal-size shares (we pad zeroes
to the secret if necessary to ensure that the CAONT package can be evenly
divided).  We encode them into $n$ shares using the systematic Reed-Solomon
codes \cite{Reed60,Blomer95,Plank97,Plank05}.


To enable deduplication, we ensure that the same share is located in the same
cloud.  Since the number of clouds for multi-cloud storage is usually small,
we simply disperse shares to all clouds.  Suppose that CDStore spans $n$
clouds, which we label $0, 1, \cdots, n-1$.
After encoding each secret using convergent dispersal, we
label the $n$ generated shares $0,1,\cdots,n-1$ in the order of their
positions in the Reed-Solomon encoding result, such that share~$i$ is
to be stored on cloud~$i$, where $0\le i\le n-1$.  This ensures that the same
cloud always receives the same share from the secrets with identical content,
either generated by the same user or different users.  This also enables us to
easily locate the shares during restore.   



\paragraph{Decoding:}
To recover the secret, we retrieve any $k$ out of $n$ shares and use
them to reconstruct the original CAONT package $(Y,t)$.  Then we deduce 
hash $h$ by XOR'ing $t$ with $\mathbf{H}(Y)$ (see
Equation~(\ref{equ:tail-part-calulation})).  Finally, we deduce secret $X$
by XOR'ing $Y$ with $\mathbf{G}(h)$ (see
Equation~(\ref{equ:main-part-calulation})), and remove any padded zeroes
introduced in encoding.

We can also verify the integrity of the deduced secret $X$.
We simply generate a hash value from the deduced $X$ as in
Equation~(\ref{equ:hash-func}) and compare if it matches $h$.  If the match
fails, then the decoded secret is considered to be corrupted.  To
obtain a correct secret, we can follow a brute-force approach, in which we try
a different subset of $k$ shares until the secret is correctly decoded
\cite{Bowers09}.

\paragraph{Remarks:} We briefly discuss the security properties of CAONT-RS.
CAONT-RS ensures confidentiality against outsider attacks, provided that an
attacker cannot gain unauthorized accesses to $k$ out of $n$ clouds, and
ensures integrity through the embedded hash in each secret.  It
leverages AONT to ensure that no information of the original secret can be
inferred from fewer than $k$ shares.  We note that an attacker can identify the
deduplication status of the shares of different users and perform brute-force
dictionary attacks \cite{Bellare13a,Bellare13b} inside the clouds, and 
we require that the secrets be drawn from a large message space (see
\S\ref{subsec:goal}).  
To mitigate brute-force attacks, we may replace the hash key in CAONT-RS with
a more sophisticated key generated by a key server \cite{Bellare13b}, with the
trade-off of introducing the key management overhead. 

\subsection{Two-Stage Deduplication}
\label{subsec:twostage}

We first overview how deduplication works.  Deduplication divides data
into fixed-size or variable-size {\em chunks}.  This work assumes
variable-size chunking, which defines boundaries based on content and is
robust to content shifting.  Each chunk is
uniquely identified by a {\em fingerprint} computed by a cryptographic hash of
the chunk content.  Two chunks are said to be identical if their fingerprints
are the same, and fingerprint collisions of two different chunks are very
unlikely in practice \cite{Black06}.  Deduplication stores only one copy of a
chunk, and refers any duplicate chunks to the copy via small-size references.  

To realize deduplication in cloud storage, a na\"{i}ve approach is to perform
global deduplication on the client side.  Specifically, before a user uploads
data to a cloud, it first generates fingerprints of the data.  It then checks
with the cloud by fingerprint for the existence of any duplicate data that
has been uploaded by any user.  Finally, it uploads only
the unique data to the cloud.  Although client-side global deduplication saves
upload bandwidth and storage overhead, it is susceptible to
{\em side-channel attacks} \cite{Harnik10,Halevi11}.  One side-channel attack
is to infer the existence of data of other users \cite{Harnik10}.
Specifically, an attacker generates the fingerprints of some possible data of
other users and queries the cloud by fingerprint if such data is unique
and needs to be uploaded. If no upload is needed, then the attacker infers 
that other users own the data.  Another side-channel attack
is to gain unauthorized access to data of other users \cite{Halevi11}.
Specifically, an attacker uses the fingerprints of some sensitive data of
other users to convince the cloud of the data ownership.

To prevent side-channel attacks, CDStore adopts \emph{two-stage
deduplication}, which eliminates duplicates first on the client side and then
on the server side.  We require that each CDStore server maintains a
deduplication index that keeps track of which shares have been stored by each
user and how shares are deduplicated (see implementation details in
\S\ref{subsec:CDStore-server-index}).  Then the two deduplication stages
are implemented as follows.


\paragraph{Intra-user deduplication:}
A CDStore client first runs
deduplication only on the data owned by the same user, and uploads the unique
data of the user to the cloud.  Before uploading shares to a cloud, the
CDStore client first checks with the CDStore server by fingerprint if it has
already uploaded the same shares.  Specifically, the CDStore client first
sends the fingerprints generated from the shares to the CDStore server.  The
CDStore server then looks up its deduplication index, and replies to the
CDStore client a list of share identifiers that indicate which shares have
been uploaded by the CDStore client.  Finally, the CDStore client uploads only
unique shares to the cloud based on the list.


\paragraph{Inter-user deduplication:}
A CDStore server runs deduplication on
the data of all users and stores the globally unique data in the cloud storage
backend.  After the CDStore server receives shares from the CDStore client, it
generates a fingerprint from each share (instead of using the one generated by
the CDStore client for intra-user deduplication), and checks if the share
has already been stored by other users by looking up the deduplication index.
It stores only the unique shares that are not yet stored at the cloud backend.
It also updates the deduplication index to keep track of which user owns the
shares.
Here, we cannot directly use the fingerprint generated by the CDStore client
for intra-user deduplication.  Otherwise, an attacker can launch a
side-channel attack, by using the fingerprint of a share of other users to
gain unauthorized access to the share \cite{Mulazzani11,Halevi11}.

\paragraph{Remarks:}
Two-stage deduplication prevents side-channel attacks by making deduplication
patterns independent across users' uploads. Thus, a malicious insider cannot
infer the data content of other users through deduplication occurrences.

Both intra-user and inter-user deduplications effectively remove duplicates.
Intra-user deduplication eliminates duplicates of the same user's data.  This
is effective for backup workloads, since the same user often makes repeated
backups of the same data as different versions \cite{Kaczmarczyk12}.
Inter-user deduplication further removes duplicates of multiple users.  For
example, multiple users within the same organization may share a large
proportion of business files.  Some workloads exhibit large
proportions of duplicates across different users' data, such as VM images
\cite{Jin09}, workstation file system snapshots \cite{Meyer11}, and backups
\cite{Wallace12}.   The removal of duplicates translates to cost savings
(see \S\ref{subsec:cost-analysis}).

\section{CDStore Implementation}
\label{sec:CDStore}

We present the implementation details of CDStore.  Our CDStore
prototype is written in C++ on Linux.  We use OpenSSL \cite{openssl} to
implement cryptographic operations: AES-256 and SHA-256 for the encryption and
hash algorithms of convergent dispersal, respectively, and SHA-256 for
fingerprints in deduplication.  We use GF-Complete \cite{Plank13d} to
accelerate Galois Field arithmetic in the Reed-Solomon coding of CAONT-RS.

\subsection{Architectural Overview}
\label{subsec:CDStore-client-architecture}

\begin{figure}[t]
\centering
\subfigure[CDStore client]{
\label{fig:CDStore-client}
\includegraphics[width=3.07in]{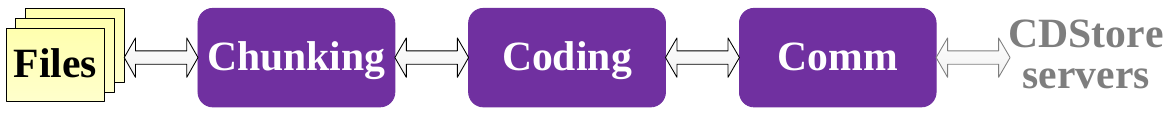}
}\\
\subfigure[CDStore server]{
\label{fig:CDStore-server}
\includegraphics[width=3.07in]{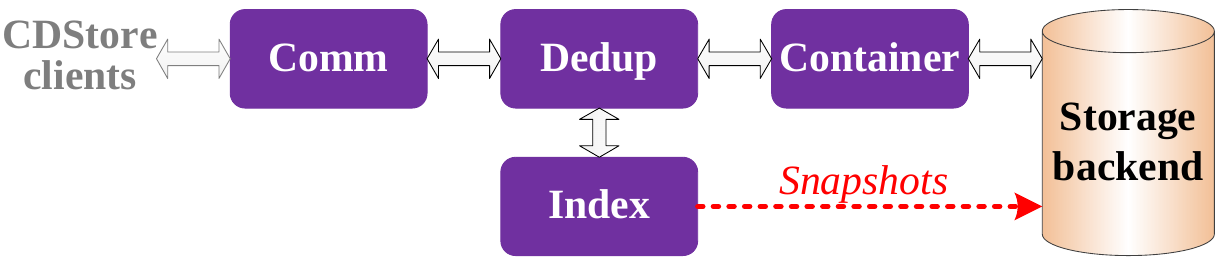}
}
\vspace{-6pt}
\caption{Implementation of the CDStore architecture.}
\label{fig:CDStore-arch}
\vspace{-6pt}
\end{figure}

We follow a modular approach to implement CDStore, whose client and server
architectures are shown in Figure~\ref{fig:CDStore-arch}.  During file
uploads, a CDStore client splits the file into a sequence of secrets via the
{\em chunking module}.
It then encodes each secret into $n$ shares via the {\em coding module}.
It performs intra-user deduplication, and uploads unique shares to
the CDStore servers in $n$ different clouds via both client-side and
server-side {\em communication modules}.
To reduce network I/Os, we avoid sending
many small-size shares over the Internet.  Instead, we first batch the shares
to be uploaded to each cloud in a 4MB buffer and upload the buffer when it is
full.  Upon receiving the shares, each CDStore server performs inter-user
deduplication via the {\em deduplication module} and updates the deduplication
metadata via the {\em index module}.  Finally, it packs the unique shares as
containers and writes the containers to the cloud storage backend through the
internal network via the {\em container module}.

File downloads work in the reverse way.  A CDStore client connects to any $k$
clouds to request to download a file.  Each CDStore server retrieves the
corresponding containers and metadata, and returns all required shares and
file metadata.  The CDStore client decodes the secrets and assembles the
secrets back to the file. 
			



\subsection{Chunking}

We implement both fixed-size chunking and variable-size chunking in the
chunking module of a CDStore client, and enable variable-size chunking by
default.  To make deduplication effective, the size of each secret should be
on the order of kilobytes (e.g., 8KB \cite{Zhu08}).  We implement variable-size
chunking based on Rabin fingerprinting \cite{Rabin81}, in which the average,
minimum, and maximum secret (chunk) sizes are configured at 8KB, 2KB, and
16KB, respectively.

\subsection{Metadata Offloading}
\label{subsec:CDStore-client-metadata}

One important reliability requirement is to tolerate client-side failures, as
we expect that a CDStore client is deployed in commodity hardware.  Thus, our
current implementation makes CDStore servers keep and manage all metadata on
behalf of CDStore clients.

When uploading a file, a CDStore client collects two types of metadata.
First, after chunking, it collects {\em file metadata} for the upload file,
including the full pathname, file size, and number of secrets.  Second, after
encoding a secret into shares, it collects {\em share metadata} for each
share, including the share size, fingerprint of the share (for intra-user
deduplication), sequence number of the input secret, and secret size (for
removing padded zeroes when decoding the original secret).

The CDStore client uploads the file and share metadata to the CDStore servers
along with the uploaded file.  The metadata will serve as input for each
CDStore server to maintain index information (see
\S\ref{subsec:CDStore-server-index}).

We distribute metadata across all CDStore servers for reliability.  For
non-sensitive information (e.g., the size and sequence number of each secret),
we can simply replicate it, so that each CDStore server can directly use it to
manage data transfer and deduplication. However, for sensitive information
(e.g., a file's full pathname), we encode and disperse it via secret sharing.

\subsection{Index Management}
\label{subsec:CDStore-server-index}

Each CDStore server uses the metadata from CDStore clients to generate index
information of the uploaded files and keep it in the index module.  There are
two types of index structures: the {\em file index} and the {\em share index}.

The file index holds the entries for all files uploaded by different users.
Each entry describes a file, identified by the full pathname (which has been
encoded as described in \S\ref{subsec:CDStore-client-metadata}) and the user
identifier provided by a CDStore client.  We hash the full pathname and the
user identifier to obtain a unique key for the entry.  The entry stores a
reference to the {\em file recipe}, which describes the complete details of
the file, including the fingerprint of each share (for retrieving the share)
and the size of the corresponding secret (for decoding the original secret).
The file recipe will be saved at the cloud backend by the container module
(see \S\ref{subsec:CDStore-server-container}).

The share index holds the entries for all unique shares of different files.
Each entry describes a share, and is keyed by the share fingerprint.  It
stores the reference to the container that holds the share.  To support
intra-user deduplication, each entry also holds a list of user identifiers to
distinguish who owns the share, as well as a reference count for each user to
support deletion.

Our prototype manages file and share indices using LevelDB
\cite{LevelDB}, an open-source key-value store.  LevelDB maintains key-value
pairs in a log-structured merge (LSM) tree \cite{ONeil96}, which supports fast
random inserts, updates, and deletes, and uses a Bloom filter \cite{Bloom70}
and a block cache to speed up lookups.  We can also leverage the snapshot
feature provided by LevelDB to store periodic snapshots in the cloud backend
for reliability. We currently do not consider this feature in our evaluation.

\subsection{Container Management}
\label{subsec:CDStore-server-container}

The container module maintains two types of containers in the storage backend:
{\em share containers}, which hold the globally unique shares, and
{\em recipe containers}, which hold the file recipes of different files.  We
cap the container size at 4MB, except that if a file recipe is very large (due
to a particularly large file), we keep the file recipe in a single
container and allow the container to go beyond 4MB.  We avoid splitting a file
recipe in multiple containers to reduce I/Os.

We make two optimizations to reduce the I/O overhead of storing and fetching
the containers via the storage backend.  First, we maintain in-memory buffers
for holding shares and file recipes before writing them into containers.  We
organize the shares or file recipes by users, so that each container contains
only the data of a single user.  This retains spatial locality of workloads
\cite{Zhu08}.  Second, we maintain a least-recently-used (LRU) disk cache to
hold the most recently accessed containers to reduce I/Os to the storage
backend.



\subsection{Multi-Threading}
\label{subsec:CDStore-multi-threading}

Advances of multi-core architectures enable us to exploit multi-threading for
parallelization.  First, the client-side coding module uses multi-threading
for the CPU-intensive encoding/decoding operations of CAONT-RS.  We
parallelize encoding/decoding at the secret level: in file uploads, we pass
each secret output from the chunking module to one of the threads for
encoding; in file downloads, we pass the shares of a secret received by the
communication module to a thread for decoding.

Furthermore, both client-side and
server-side communication modules use multi-threading to fully utilize the
network transfer bandwidth.  The client-side communication module creates
multiple threads, one for each cloud, to upload/download shares.  The
server-side communication module also uses multiple threads to send/receive
shares for different CDStore clients.


\subsection{Open Issues}

Our current CDStore prototype implements the basic backup and restore
operations.  We discuss some open implementation issues.

\paragraph{Storage efficiency:}  We can reclaim more storage space via
different techniques in addition to deduplication.  For example, garbage
collection can reclaim space of expired backups.  By exploiting historical
information, we can accelerate garbage collection in deduplication storage
\cite{Fu14}.  Compression also effectively reduces storage space of both
data \cite{Wallace12} and metadata (e.g., file recipes \cite{Meister13}).
Implementations of garbage collection and compression are posed as future
work. 


\paragraph{Scalability:}  We currently deploy one CDStore server per cloud.
In large-scale deployment, we can run CDStore servers on multiple VMs per
cloud and evenly distribute user backup jobs among them for load balance.
Implementing a distributed deduplication system is beyond the scope of this
paper.  


\paragraph{Consistency:}  Our prototype is tailored for backup workloads that
are immutable.  We do not address consistency issues due to concurrent updates
as mentioned in \cite{Bessani13}.

\section{Evaluation}
\label{sec:evaluation}

We evaluate CDStore under different testbeds and workloads. We also analyze
its monetary cost advantages. 

\subsection{Testbeds}
\label{subsec:testbeds}

We consider three types of testbeds in our evaluation.

{\em (i) Local machines:}  We use two machines: {\em Local-Xeon}, which 
has a quad-core 2.4GHz Intel Xeon E5530 and 16GB RAM, and {\em Local-i5},
which has a quad-core 3.4GHz Intel Core i5-3570 and 8GB RAM.   Both
machines run 64-bit Ubuntu 12.04.2 LTS.
We use them to evaluate the encoding performance of CDStore clients.

{\em (ii) LAN:}  We configure a LAN of multiple machines with the same
configuration as Local-i5.  All nodes are connected via a 1Gb/s switch.  We
run CDStore clients and servers on different machines.  Each CDStore server
mounts the storage backend on a local 7200RPM SATA hard disk.  We use the LAN
testbed to evaluate the data transfer performance of CDStore. 

{\em (iii) Cloud:}  We deploy a CDStore client on the Local-Xeon machine (in
Hong Kong) and connect it via the Internet to four commercial clouds (i.e.,
$n=4$): Amazon (in Singapore), Google (in Singapore), Azure (in Hong Kong),
and Rackspace (in Hong Kong).  We set up the testbed in the same continent to
limit the differences among the client-to-server connection bandwidths.
Each cloud runs a VM with similar configurations: four CPU cores and
4$\sim$15GB RAM.  We use the cloud testbed to evaluate the real deployment
performance of CDStore. 

\subsection{Datasets}

We use two real-world datasets to drive our evaluation. 

{\em (i) FSL:}  This dataset is published by the File systems and Storage Lab
(FSL) at Stony Brook University \cite{FSL,Tarasov12}.  Due to the large
dataset size, we use the {\tt Fslhomes} dataset in 2013, containing daily
snapshots of nine students' home directories from a shared network file
system.  We select the snapshots every seven days (which are not continuous)
to mimic weekly backups.
The dataset is represented in 48-bit chunk fingerprints and corresponding
chunk sizes obtained from variable-size chunking.  Our filtered FSL dataset
contains 16 weekly backups of all nine users, covering a total of 8.11TB of
data.


{\em (ii) VM:} This dataset is collected by ourselves and is unpublished. It
consists of weekly snapshots of 156 VM images for students in a university
programming course in Spring 2014.  We create a 10GB master image with Ubuntu
12.04.2 LTS and clone all VMs.  We treat each VM image snapshot as a weekly
backup of a user.  The dataset is represented in SHA-1 fingerprints on 4KB
fixed-size chunks.  It spans 16 weeks, totaling 24.38TB of data.  For fair
comparisons, we remove all zero-filled chunks (which dominate in VM images
\cite{Jin09}) from the dataset, and the size reduces to 11.12TB.


\subsection{Encoding Performance}
\label{subsec:encoding-speed}

We evaluate the computational overhead of CAONT-RS when encoding secrets into
shares.  We compare CAONT-RS with two variants: (i) {\em AONT-RS}
\cite{Resch11}, which builds on Rivest's AONT \cite{Rivest97} and does not
support deduplication, and (ii) our prior proposal {\em CAONT-RS-Rivest}
\cite{Li14b}, which uses Rivest's AONT as in AONT-RS and replaces the random
key in AONT-RS with a SHA-256 hash for convergent dispersal.  CAONT-RS uses 
OAEP-based AONT instead (see \S\ref{subsec:CD}).

We conduct our experiments on the Local-Xeon and Local-i5 machines.  We create
2GB of random data in memory (to remove I/O overhead), generate secrets using
variable-size chunking with an average chunk size 8KB, and encode them into
shares.  We measure the {\em encoding speed}, defined as the ratio of the
original data size to the total time of encoding all secrets into shares.  Our
results are averaged over 10 runs.  We observe similar results for decoding,
and omit them here.

\begin{figure}[t]
  \centering
  \subfigure[Varying number of threads]{
    \label{fig:encoding_speed_thread}
    \includegraphics[width=3in]{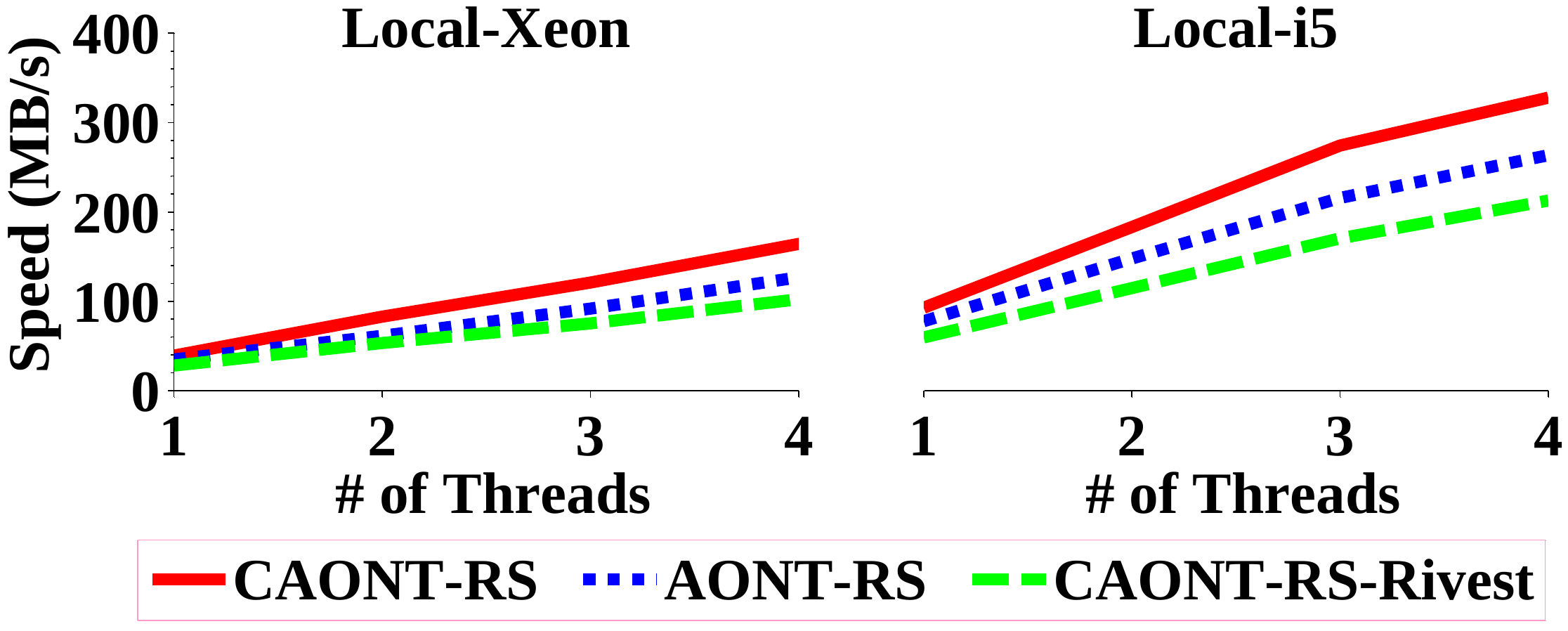}
  }
  \hspace{1in}
  \subfigure[Varying $n$]{
    \label{fig:encoding_speed_n}
    \includegraphics[width=3in]{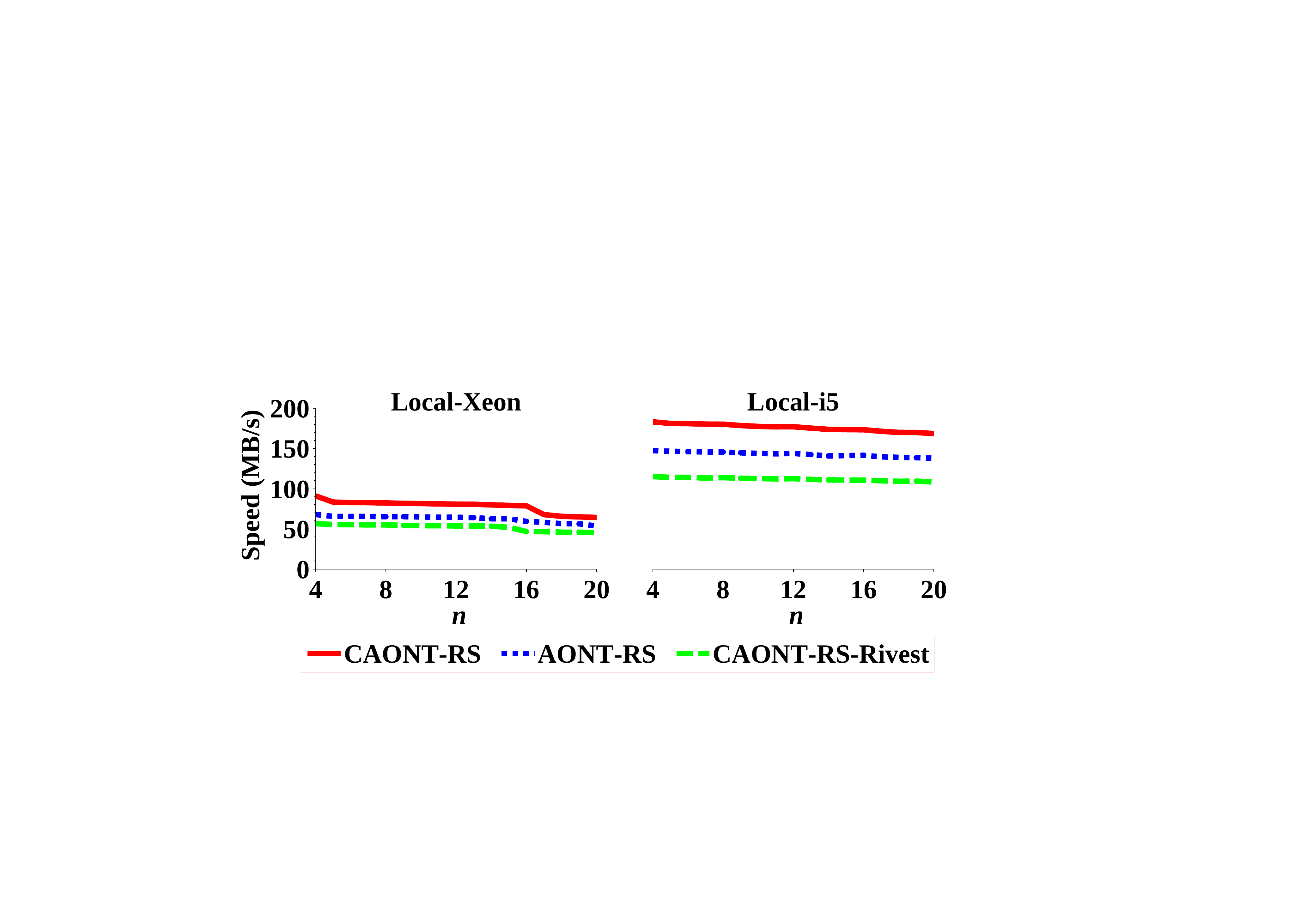}
  }
  \vspace{-1em}
  \caption{Encoding speeds of a CDStore client.}
  \label{fig:encoding_speed}
\end{figure}

We first examine the benefits of multi-threading (see
\S\ref{subsec:CDStore-multi-threading}).
Figure~\ref{fig:encoding_speed_thread} shows the encoding speeds versus the
number of threads, while we fix $(n,k) = (4,3)$.  The encoding speeds of all
schemes increase with the number of threads.  If two encoding threads are
used, the encoding speeds of CAONT-RS are 83MB/s on Local-Xeon and 183MB/s on
Local-i5. Also, OAEP-based AONT in CAONT-RS brings remarkable performance
gains.  Compared to CAONT-RS-Rivest, which performs encryptions on small words
based on Rivest's AONT, CAONT-RS improves the encoding speed by 40$\sim$61\%
on Local-Xeon and 54$\sim$61\% on Local-i5; even though compared to AONT-RS,
which uses one fewer hash operation, CAONT-RS still increases the encoding
speed by 12$\sim$35\% on Local-Xeon and 19$\sim$27\% on Local-i5.


We next evaluate the impact of $n$ (number of clouds). We vary $n$ from 4 to
20, and fix two encoding threads.  We configure $k$ as the largest integer
that satisfies $\frac{k}{n} \leq \frac{3}{4}$ (e.g., $n=4$ implies $k=3$), so
as to maintain a similar storage blowup due to secret sharing.
Figure~\ref{fig:encoding_speed_n} shows the encoding speeds versus $n$.  The
encoding speeds of all schemes slightly decrease with $n$ (e.g., by 8\%
from $n=4$ to $20$ for CAONT-RS on Local-i5), since more encoded shares are
generated via Reed-Solomon codes for a larger $n$.  However, Reed-Solomon
coding only accounts for small overhead compared to AONT, which runs
cryptographic operations.  We have also tested other ratios of $\frac{k}{n}$
and obtained similar speed results.

The above results only report encoding speeds, while a CDStore client performs
both chunking and encoding operations when uploading data to multiple clouds.
We measure the combined chunking (using variable-size chunking) and encoding
speeds with $(n,k)=(4,3)$ and two encoding threads, and find that the combined
speeds drop by around 16\%, to 69MB/s on Local-Xeon and 154MB/s on Local-i5.



\subsection{Deduplication Efficiency}
\label{subsec:trace-efficiency}

\begin{figure}[t]
  \centering
  \subfigure[Intra-user and inter-user deduplication savings]{
    \label{fig:dedup_saving_combined}
    \includegraphics[width=3in]{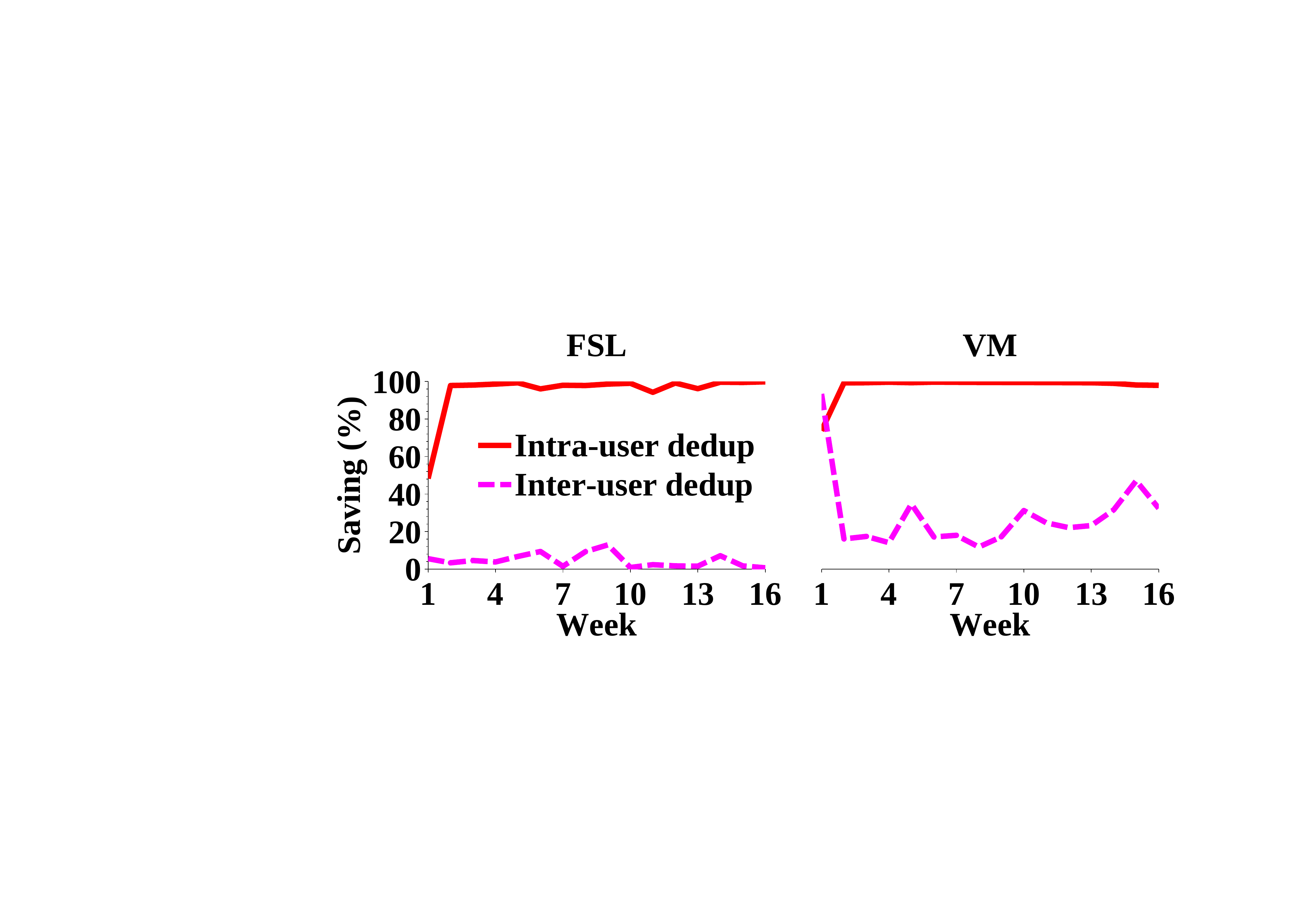}
  }
  \subfigure[Cumulative data and share sizes when $(n,k)=(4,3)$]{
    \label{fig:cumulative_size_combined}
    \includegraphics[width=3in]{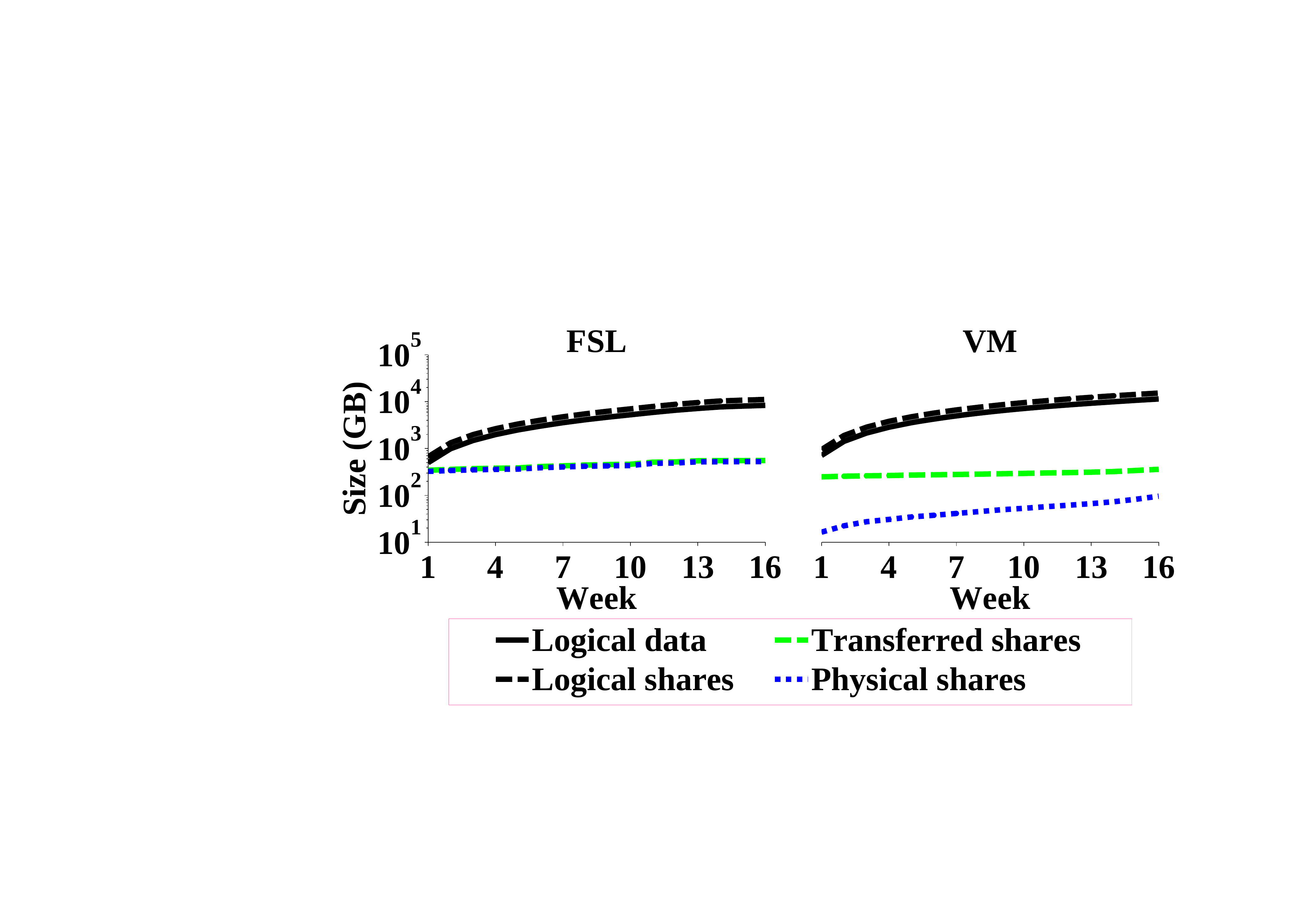}
  }
  \vspace{-1em}
  \caption{Deduplication efficiency of CDStore.}
  \label{fig:dedup_efficiency}
\end{figure}

We evaluate the effectiveness of both intra-user and inter-user deduplications
(see \S\ref{subsec:twostage}).  We extract the deduplication characteristics
of both datasets, assuming that they are stored as weekly
backups.  We define four types of data: (i) \emph{logical data}, the original
user data to be encoded into shares, (ii) \emph{logical shares}, the shares
before two-stage deduplication, (iii) \emph{transferred shares}, the shares
that are transferred over Internet after intra-user deduplication, and 
(iv) \emph{physical shares}, the shares that are finally stored after 
two-stage deduplication.  We also define two metrics: 
(i) \emph{intra-user deduplication saving}, which is one minus the
ratio of the size of the transferred shares to that of the
logical shares, and (ii) \emph{inter-user deduplication saving}, which is one
minus the ratio of the size of the physical shares to that of the transferred
shares.  We fix $(n,k)=(4,3)$.
Figure~\ref{fig:dedup_efficiency} summarizes the results.


Figure~\ref{fig:dedup_saving_combined} first shows the intra-user and 
inter-user deduplication savings.  The intra-user deduplication savings 
are very high for both datasets, especially in subsequent backups 
after the first week (at least 94.2\% for FSL and at least 98.0\% for VM). 
The reason is that the users only modify or add a small portion of data.  
The savings translate to performance gains in file uploads 
(see \S\ref{subsec:trace-evaluation}). 
However, the inter-user deduplication savings differ across datasets.  
For the FSL dataset, the savings fall to no more than 12.9\%.  
In contrast, for the VM dataset, the saving for the first
backup reaches 93.4\%, mainly because the VM images are initially installed
with the same operating system.  The savings for subsequent backups then drop
to the range between 11.8\% and 47.0\%.  Nevertheless, the VM dataset shows
higher savings for subsequent backups than the FSL dataset; we conjecture the
reason is that students make similar changes to the VM images when doing
programming assignments.

Figure~\ref{fig:cumulative_size_combined} then shows cumulative data and 
share sizes before and after intra-user and inter-user deduplications. 
After 16 weekly backups, for the FSL dataset, the
total size of physical shares is only 0.51TB, about 6.3\% of the logical data
size; for the VM dataset, the total size of physical shares is only 0.09TB,
about 0.8\% of the logical data size.  This shows that dispersal-level
redundancy (i.e., $\frac{n}{k} = \frac{4}{3}$) is significantly offset by
removing content-level redundancy via two-stage deduplication.  Also, if we
compare the sizes of transferred shares and physical shares for the VM
dataset, we see that inter-user deduplication is crucial for reducing storage
space.

\subsection{Transfer Speeds}
\label{subsec:trace-evaluation}


\begin{figure*}[!t]
\centering
\begin{tabular}{cc}
\parbox[t]{4.2in}{
\subfigure[Baseline results]{
\label{fig:client_speed}
\hspace{-0.1in}
\includegraphics[width=2.15in]{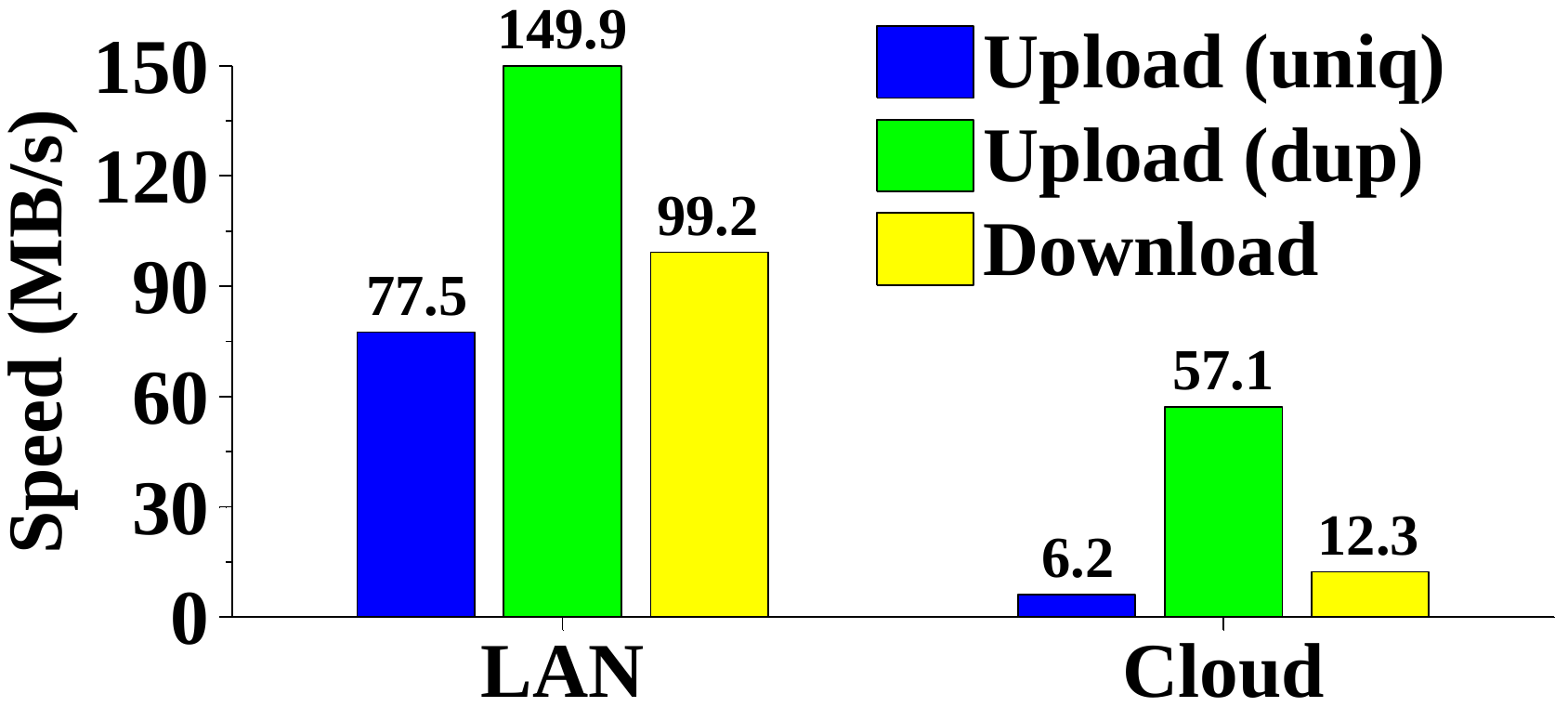}
}
\subfigure[Trace-driven results]{
\label{fig:trace_speed}
\hspace{-0.1in}
\includegraphics[width=2.15in]{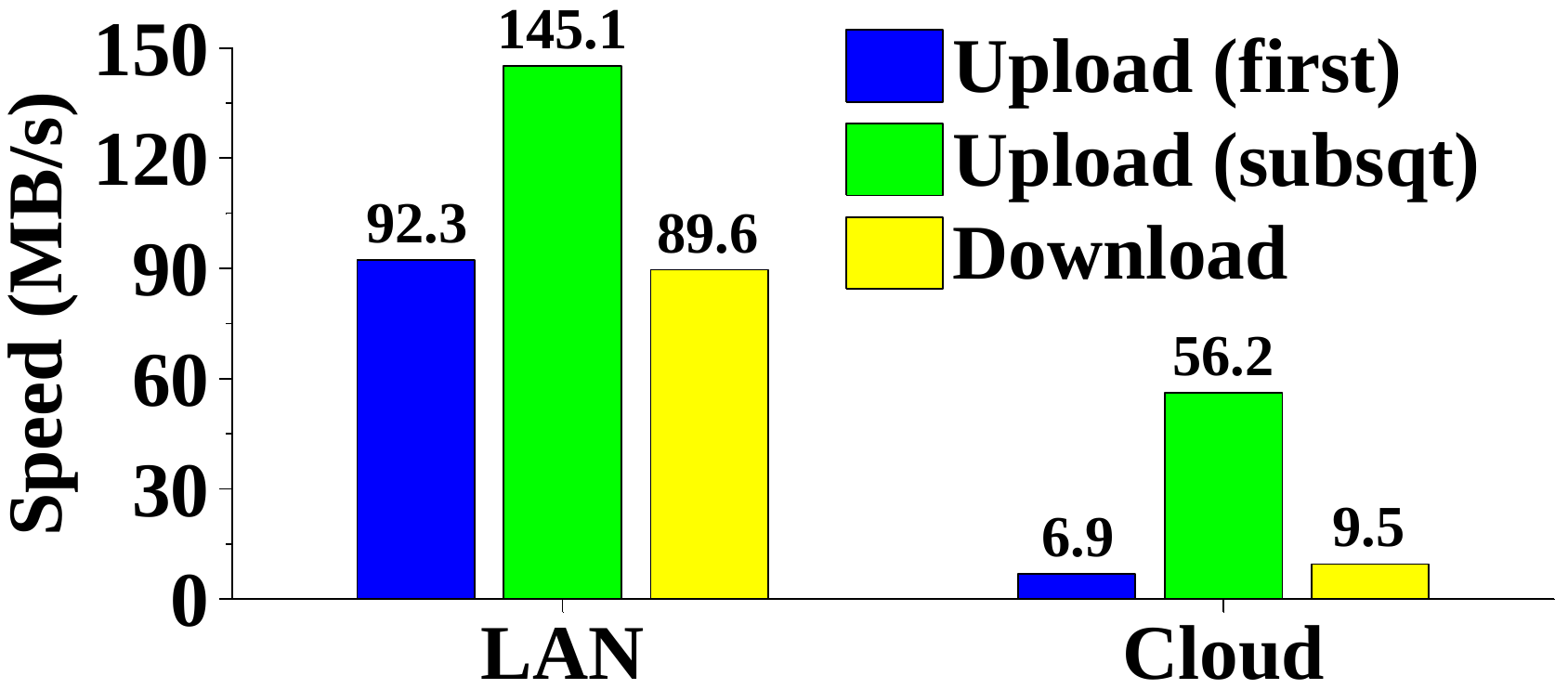}
}
\vspace{-1em}
\caption{Upload and download speeds of a CDStore client (the numbers are
the speeds in MB/s).}
\label{fig:test_speed}
}
&
\parbox[t]{2in}{
\vspace{0.1in}
\includegraphics[width=2.1in]{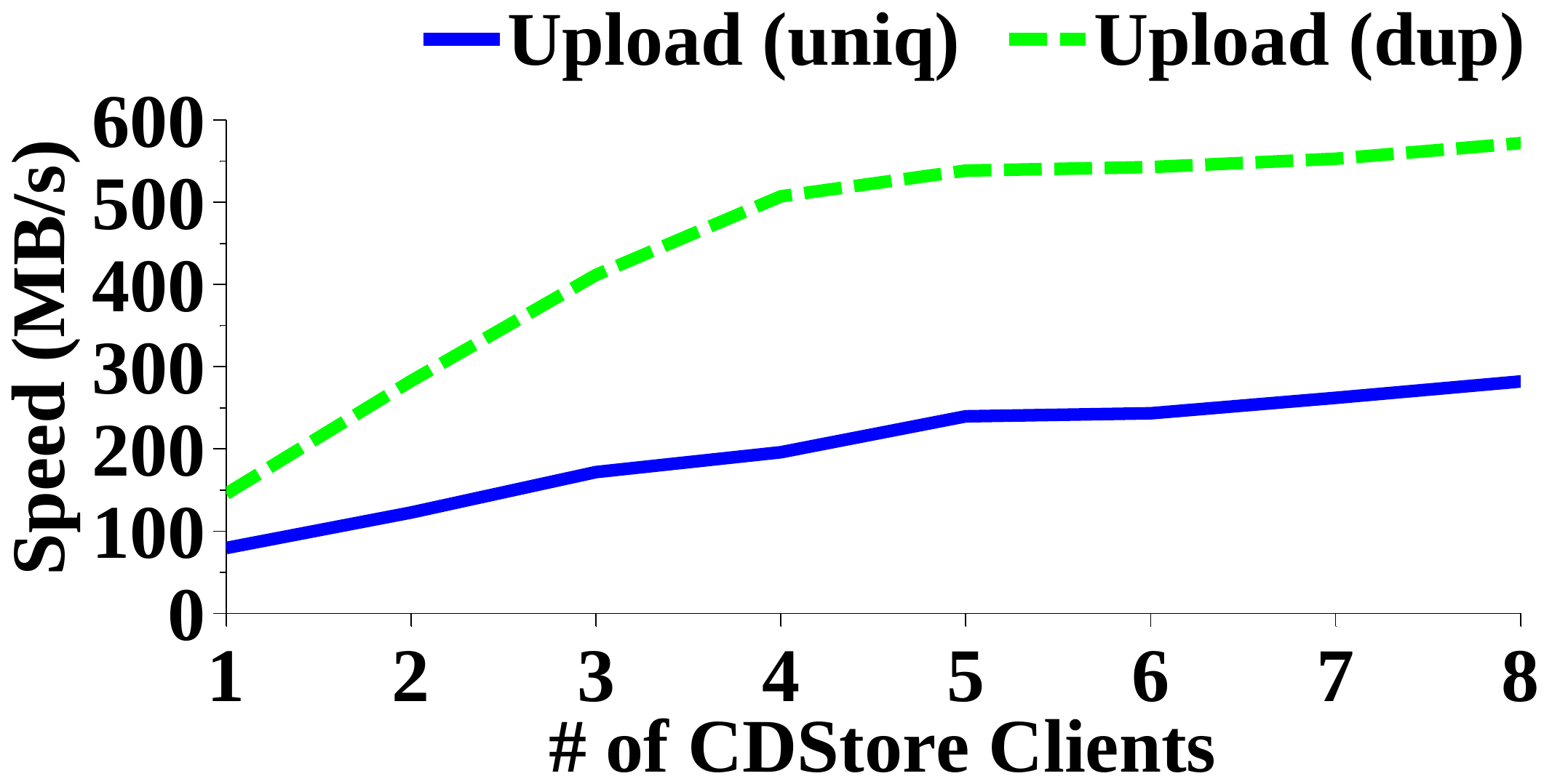}
\caption{Aggregate upload speeds of multiple CDStore clients.}
\label{fig:server_speed}
}
\end{tabular}
\vspace{-6pt}
\end{figure*}

\paragraph{Single-client baseline transfer speeds:}
We first evaluate the baseline transfer speed of a CDStore client using
both LAN and cloud testbeds.  Each testbed has one CDStore client and four
CDStore servers with $(n,k)=(4,3)$.  We first upload 2GB of unique data (i.e.,
no duplicates), then upload another 2GB of duplicate data identical to the
previous one, and finally download the 2GB data from three CDStore servers
(for the cloud testbed, we choose Google, Azure, and Rackspace for downloads).
We measure the upload and download speeds, averaged over 10 runs.




Figure~\ref{fig:client_speed} presents the results.  On the LAN testbed, the
upload speed for unique data is 77MB/s.  Our measurements find that the
effective network speed in our LAN testbed is around 110MB/s.  Thus, the
upload speed for unique data is close to $\frac{k}{n}$ of the effective
network speed.  Uploading duplicate data has speed 150MB/s.  Since it does not
transfer actual data after intra-user deduplication, the performance is
bounded by the chunking and CAONT-RS encoding operations (see
\S\ref{subsec:encoding-speed}).  The download speed is 99MB/s, about 10\%
less than the effective network speed.  The reason is that
the CDStore servers need to retrieve data from the disk backend before
returning it to the CDStore client.


On the cloud testbed, the upload and download performance is limited by the
Internet bandwidth.  For references, we measure the upload and download speeds
of each individual cloud when transferring 2GB of unique data divided in 4MB
units (see \S\ref{subsec:CDStore-client-architecture}), and
Table~\ref{tab:cloud_speed} presents the averaged results over 10 runs.  Since
CDStore transfers data through multiple clouds in parallel via
multi-threading, its upload speed of unique data and download speed are higher
than those of individual clouds (e.g., Amazon and Google).  The upload speed
for unique data is smaller than the download speed because of sending
redundancy and connecting to more clouds.  The upload speed for duplicate data
is over 9$\times$ that for unique data, and this difference is more
significant than on the LAN testbed.  



\begin{table}[t]
\centering
\small
\begin{tabular}{ccc}
\toprule
\textbf{Cloud}  & \textbf{Upload speed} & \textbf{Download speed} \\
\midrule
Amazon & 5.87 (0.19) & 4.45 (0.30) \\
Google & 4.99 (0.23) & 4.45 (0.21) \\
Azure & 19.59 (1.20) & 13.78 (0.72) \\
Rackspace & 19.42 (1.06) & 12.93 (1.47) \\
\bottomrule
\end{tabular}
\caption{Measured speeds (MB/s) of each of four clouds, in terms of the
average (standard deviation) over 10 runs.}
\label{tab:cloud_speed}
\vspace{-3pt}
\end{table}

\paragraph{Single-client trace-driven transfer speeds:}
We now evaluate the upload and download
speeds of a single CDStore client using datasets as opposed to unique and
duplicate data above.  We focus on the FSL dataset, which allows us to test
the effect of variable-size chunking.  We again consider both LAN and cloud
testbeds with $(n,k)=(4,3)$.  Since the FSL dataset only has chunk
fingerprints and chunk sizes, we reconstruct a chunk by writing the
fingerprint value repeatedly to a chunk with the specified size, so as to
preserve content similarity.  
Each chunk is treated as a secret, which will be encoded into shares.
We first upload all backups to CDStore servers, followed by downloading them.
To reduce evaluation time, we only run part of the dataset.  On the LAN
testbed, we run seven weekly backups for five users (1.06TB data in total). We
feed the first week of backups of each user one by one through the CDStore
client, followed by the second week of backups, and so on.  
On the other hand, on the cloud testbed, we run two weekly backups for a
single user (21.35GB data in total).  

Figure~\ref{fig:trace_speed} presents three results:
(i) the average upload speed for the first backup (averaged over five users
for the LAN testbed), (ii) the average upload speed for the subsequent
backups, and (iii) the average download speed of all backups.  The presented
results are obtained from a single run, yet the evaluation time is long enough
to give steady-state results.  We compare the results with those for unique and
duplicate data in Figure~\ref{fig:client_speed}.

We see that the upload speed for the first backup exceeds that for unique data
(e.g., by 19\% on the LAN testbed), mainly because the first
backup contains duplicates, which can be removed by intra-user deduplication
(see Figure~\ref{fig:dedup_saving_combined}).  The upload speed for
the subsequent backups approximates to that for duplicate data, as most
duplicates are again removed by intra-user deduplication.

The trace-driven download speed is lower than the baseline
one in Figure~\ref{fig:client_speed} (e.g., by 10\% on the LAN testbed), since
deduplication now introduces chunk fragmentation \cite{Lillibridge13} for
subsequent backups.  Nevertheless, we find that the
variance of the download speeds of the backups is very small (not shown in the
figure), although the number of accessed containers increases for subsequent
backups.   The download speed will gradually degrade due to fragmentation as
we store more backups.  We do not explicitly address fragmentation in this
work.


\paragraph{Multi-client aggregate upload speeds:}
We evaluate the aggregate upload speed
when multiple CDStore clients connect to multiple CDStore servers.  We
mainly consider data uploads on the LAN testbed, in which we vary the number of
CDStore clients, each hosted on a dedicated machine, and 
configure four CDStore servers with $(n,k)=(4,3)$.  All CDStore clients
perform uploads concurrently, such that each of them first uploads 2GB of
unique data, and then uploads another 2GB of duplicate data.  We measure the 
{\em aggregate upload speed}, defined as the
total upload size (i.e., 2GB times the number of clients) divided by the
duration when all clients finish uploads.  Our results are averaged over 10
runs.

Figure~\ref{fig:server_speed} presents the aggregate upload speeds for both
unique and duplicate data, which we observe increase with the number of
CDStore clients.  For unique data, the aggregate upload speed reaches 282MB/s
for eight CDStore clients.  The speed is limited by the network bandwidth and
disk I/O, where the latter is for the CDStore servers to write containers to
disk.  If we exclude disk I/O (i.e., without writing data), the aggregate
upload speed can reach 310MB/s (not shown in the figure), which approximates
to the aggregate effective Ethernet speed of $k=3$ CDStore servers.  For
duplicate data, there is no actual data transfer, so the aggregate upload
speed can reach 572MB/s.  Note that the knee point at four CDStore clients is
due to the saturation of CPU resources in each CDStore server.

\subsection{Cost Analysis}
\label{subsec:cost-analysis}


We now analyze the cost saving of CDStore.  We compare it with two baseline
systems: (i) an AONT-RS-based multi-cloud system that has the same levels of
reliability and security as CDStore but does not support deduplication, and
(ii) a single-cloud system that incurs zero redundancy for reliability, but
encrypts user data with random keys and does not support deduplication.  We
aim to show that CDStore incurs less cost than AONT-RS through deduplication;
even though CDStore incurs redundancy for reliability, it still incurs less cost than
the single-cloud system without deduplication.


We develop a tool to estimate the monetary costs using the pricing models of
Amazon EC2 \cite{AmazonEC2} and S3 \cite{AmazonS3} in September 2014.  Free
charges apply to data transfers between co-locating EC2 instances and S3
storage, and also inbound transfers to both EC2 and S3.  We only study backup
operations, and do not consider restore operations as they are relatively
infrequent in practice.  Note that both EC2 and S3 follow tiered pricing, so
the exact charges depend on the actual usage.  Our tool takes into account
tiered pricing in cost calculations. For CDStore, we also consider the storage
costs of file recipes.

We briefly describe how we derive the EC2 and S3 costs.  For EC2, we consider
the category of high-utilization reserved instances, which are priced based on
an upfront fee and hourly bills.  We focus on two types of instances, namely
compute-optimized and storage-optimized, to host CDStore servers on all
clouds.  Each instance charges around US\$60$\sim$1,300 per month,
depending on the CPU, memory, and storage settings.  Note that both file and
share indices (see \S\ref{subsec:CDStore-server-index}) are kept in the
local storage of an EC2
instance, and the total index size is determined by how much data is stored
and how much data can be deduplicated.  Our tool chooses the cheapest instance
that can keep the entire indices according to the storage size and
deduplication efficiency, both of which can be estimated in practice.
On the other hand, S3 storage is mainly priced based on storage size, and it
charges around US\$30 per TB per month.   Note that in backup operations, the
costs due to outbound transfer (e.g., a CDStore server replies the intra-user
deduplication status to a CDStore client) and storage requests (e.g., PUT) are
negligible compared to VM and storage costs.

We consider a case study. An organization schedules weekly backups for its user
data, for a retention time of half a year (26 weeks).   We fix $(n,k)=(4,3)$
(i.e., we host four EC2 instances for CDStore servers).  We vary the weekly
backup size and the deduplication ratio, where the latter is defined as the
ratio of the size of logical shares to the size of physical shares (see
\S\ref{subsec:trace-efficiency}).

\begin{figure}[t]
  \centering
  \subfigure[Varying weekly backup size]{
    \label{fig:cost_analysis_week_size}
    \includegraphics[width=1.47in]{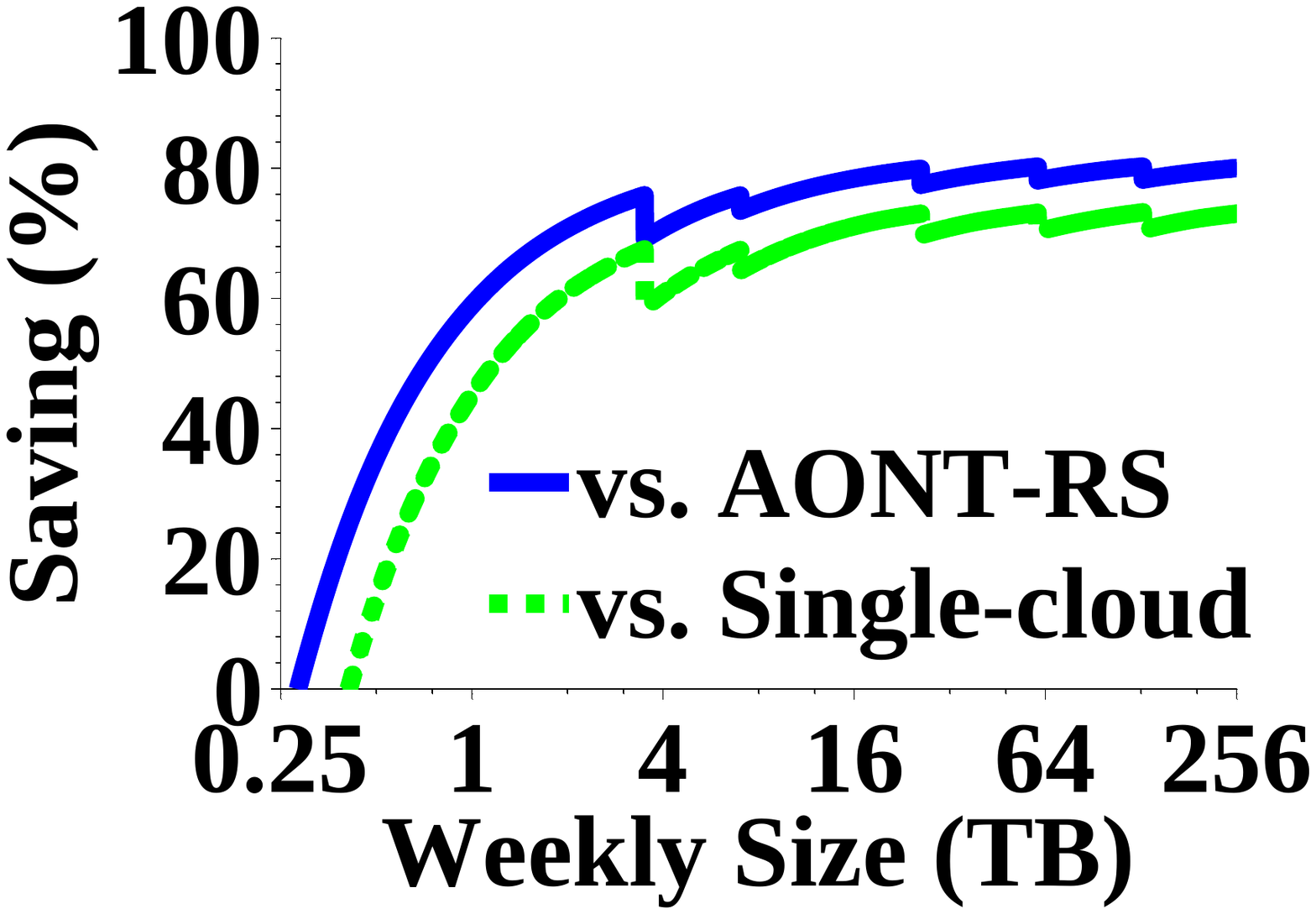}
  }
  \subfigure[Varying deduplication ratio]{
    \label{fig:cost_analysis_dedup_ratio}
    \includegraphics[width=1.47in]{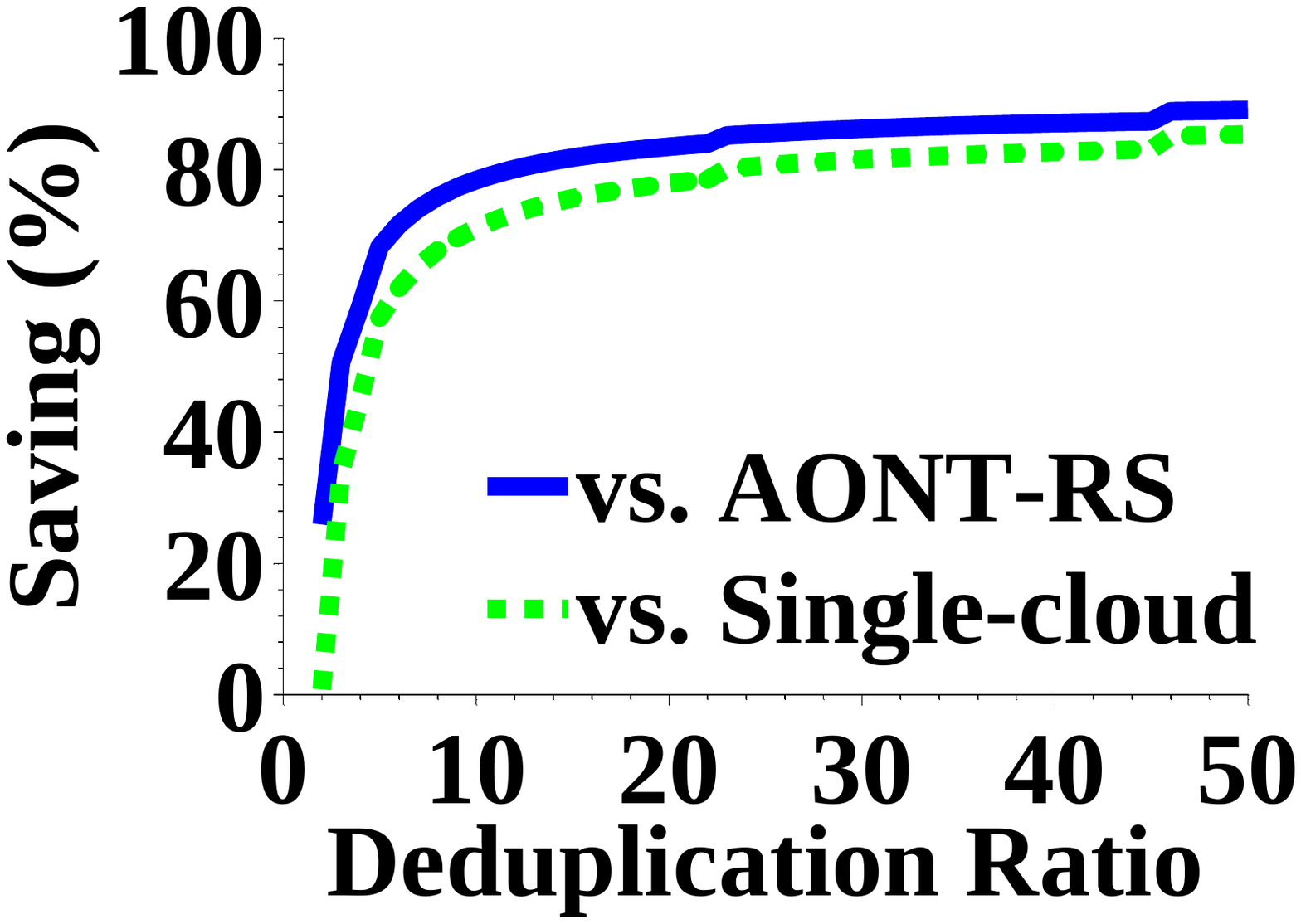}
  }
  \vspace{-6pt}
  \caption{Cost savings of CDStore over an AONT-RS-based multi-cloud system and a single-cloud system.}
  \label{fig:cost_analysis}
\end{figure}

Figure~\ref{fig:cost_analysis_week_size} shows the cost savings of
CDStore versus different weekly backup sizes, while we fix the deduplication
ratio as 10$\times$ \cite{Wallace12}.  The cost savings increase with the
weekly backup size.
For example, if we keep a weekly backup size of 16TB,
the single-cloud and AONT-RS-based systems incur total storage costs (with tiered
pricing) of around US\$12,250/month and US\$16,400/month, respectively; CDStore
incurs additional VM costs of around US\$660/month but reduces the storage cost to
around US\$2,880/month, resulting in around US\$3,540/month in total and thus
achieving at least 70\% of cost savings as a whole.
The cost saving of CDStore over AONT-RS is higher than that over a single
cloud, as the former introduces dispersal-level redundancy for fault
tolerance.
The increase slows down as the weekly backup size further increases, since the
overhead of file recipes becomes significant when the total backup size is
large while the backups have a high deduplication ratio \cite{Meister13}.
Note that the jagged curves are due to the switch of the cheapest EC2
instance to fit the indices.

Figure~\ref{fig:cost_analysis_dedup_ratio} shows the cost savings of CDStore
versus different deduplication ratios, where the weekly backup size is fixed
at 16TB.  The cost saving increases with the deduplication ratio.  The saving
is about 70$\sim$80\% when the deduplication ratio is between 10$\times$ and
50$\times$.

\section{Related Work}
\label{sec:related}



\paragraph{Multi-cloud storage:} Existing multi-cloud storage systems mainly
focus on data availability in the presence of cloud failures and vendor
lock-ins.  For example, SafeStore \cite{Kotla07}, RACS \cite{Abu-Libdeh10},
Scalia \cite{Papaioannou12}, and NCCloud \cite{Hu12} disperse redundancy
across multiple clouds using RAID or erasure coding. 
Some multi-cloud systems additionally address security.  HAIL \cite{Bowers09}
proposes proof of retrievability to support remote integrity checking against
data corruptions.  MetaStorage \cite{Bermbach11} and SPANStore \cite{Wu13}
provide both availability and integrity guarantees by replicating data across
multiple clouds using quorum techniques \cite{Malkhi97}, but do not
address confidentiality.   Hybris \cite{Dobre14} achieves confidentiality by
dispersing encrypted data over multiple public clouds via erasure coding and
keeping secret keys in a private cloud.  


\paragraph{Applications of secret sharing:}  We discuss several secret sharing
algorithms in \S\ref{sec:background}. They have been realized by storage
systems.  POTSHARDS \cite{Storer09} realizes Shamir's scheme \cite{Shamir79}
for archival storage.  ICStore \cite{Cachin10} achieves
confidentiality via key-based encryption, where the keys are
distributed across multiple clouds via Shamir's scheme.  DepSky
\cite{Bessani13} and SCFS \cite{Bessani14} distribute keys across clouds using
SSMS \cite{Krawczyk93}.  Cleversafe \cite{Resch11} uses AONT-RS to achieve
security with reduced storage space.  All the above systems rely on random
inputs to secret sharing, and do not address deduplication.  


\paragraph{Deduplication security:} Convergent encryption
\cite{Douceur02} provides confidentiality guarantees for deduplication
storage, and has been adopted in various storage systems
\cite{Adya02,Cox02,Storer08,Wilcox08,Anderson10}.  However, the key
management overheads of convergent encryption are significant \cite{Li14a}.
Bellare {\em et al.} \cite{Bellare13a} generalize convergent encryption into
Message-locked encryption (MLE) and provide formal security analysis on
confidentiality and tag consistency.  The same authors also prototype a
server-aided MLE system DupLESS \cite{Bellare13b}, which uses more complicated
encryption keys to prevent brute-force attacks.  DupLESS maintains the keys in
a dedicated key server, yet the key server is a single point of failure.

Client-side inter-user deduplication poses
new security threats, including the side-channel attack
\cite{Harnik10,Halevi11} and some specific attacks against Dropbox
\cite{Mulazzani11}.  CDStore addresses this problem through two-stage
deduplication.  A previous work \cite{Zhang12} proposes a similar two-stage
deduplication approach (i.e., inner-VM and cross-VM deduplications) to reduce
system resources for VM backups, while our approach is mainly to address
security.

\section{Conclusions}
\label{sec:conclusions}

We propose a multi-cloud storage system called CDStore for 
organizations to outsource backup and archival storage to
public cloud vendors, with three goals in mind: reliability,
security, and cost efficiency.  The core design of CDStore is convergent
dispersal, which augments secret sharing with the deduplication capability.
CDStore also adopts two-stage deduplication to achieve bandwidth and storage
savings and prevent side-channel attacks.  We extensively evaluate CDStore via
different testbeds and datasets from both performance and cost perspectives.
We demonstrate that deduplication enables CDStore to achieve cost savings.
The source code of our CDStore prototype is available at 
{\bf http://ansrlab.cse.cuhk.edu.hk/software/cdstore}.

\section*{Acknowledgments}

We would like to thank our shepherd, Fred Douglis, and the anonymous reviewers
for their valuable comments.  This work was supported in part by grants 
ECS CUHK419212 and GRF CUHK413813 from HKRGC.

\bibliographystyle{abbrv}
\bibliography{reference}



\end{document}